\documentclass[12pt]{article}

\author{Yu.~M.~Zinoviev
       \thanks{E-mail address: ZINOVIEV@MX.IHEP.SU} \\
        {\it Institute for High Energy Physics} \\
        {\it Protvino, Moscow Region, 142280, Russia}}
\title{First Order Formalism \\
       for Massive Mixed Symmetry Tensor Fields \\
		in Minkowski and (A)dS Spaces}

\date{}

\begin{document}
\maketitle

\begin{abstract}
In this paper we extend our recent results (hep-th/0304067) on
the first order formulation for the massless mixed symmetry tensor
fields to the case of massive fields both in Minkowski as well as
in (Anti) de Sitter spaces (including all possible massless and
partially massless limits). Main physical results are essentially
the same as in hep-th/0211233.
\end{abstract}

\thispagestyle{empty}

\newpage
\setcounter{page}{1}

\section*{Introduction}

Last times there is a renewed interest in the mixed symmetry high
spin tensor fields \cite{Cur86}-\cite{BPT01}. The reason is that such
fields naturally appear in a number of physically interesting theories
such as superstrings, higher dimensional supergravities and 
(supersymmetric) high spin theories. One of the technical difficulties
(besides purely combinatorial ones) one faces working with such
fields is that to get analog of gauge invariant "field strengths" one 
has to build expressions with more and more derivatives (or has to 
work with non-local terms in the equations of motion or the 
Lagrangians)\cite{BB02,MH02,BB03,MH03}. The structure of gauge 
transformation laws is also appears to be rather complicated, moreover, 
these transformations often turn out to be reducible. All this make the
problem of investigation of possible interactions among such fields
a very complicated task 
\cite{BG00,BBH02,BCNS02,BCCSS03,Sal03,BCCSS03a}.
Recently \cite{Zin03} we have constructed a few examples of
first order formulation for such mixed symmetry fields which turned
out to be very similar to the well known tetrad formulation of
gravity\footnote{First order formulation for high spin fields
corresponding to symmetric tensors has been given recently in
\cite{Vas03}}. The Lagrangians obtained have simple and very
suggestive form, so that one could hope that such formulation can
help in investigations of possible interactions among these fields.
Let us note also that first order "parent" Lagrangians play a very
important role in investigations of dualities for such fields
\cite{CMU03,BCH03}. In this paper we construct examples of first order
formulation for massive mixed symmetry tensor fields both in
flat Minkowski as well as in (Anti) de Sitter spaces. Particles
in (A)dS reveal a number of very peculiar features such as unitary
forbidden regions (i.e. not all values of mass and cosmological
constant are allowed) and appearance of partially massless theories
\cite{DN83}-\cite{Tek03}. Moreover not all massless fields in flat
Minkowski space could be deformed into the (A)dS space without
introduction of additional fields \cite{BMV00}. In our previous
works on this subject \cite{Zin01,Zin02} we used gauge invariant
description of massive particles. Such description being gauge
invariant and unitary from the very beginning turns out to be very
well suited for the investigation of unitarity, gauge invariance 
and partial masslessness. In this paper we construct such gauge
invariant description of massive particles using a first order
formalism for all appropriate massless components.

The paper is organized as follows. In the following section we
give simple but very instructive example of massive spin-2
particle. In the next section we give a first order gauge invariant
formulation for the simplest mixed symmetry tensor field
$\Phi_{\mu\nu}{}^a$ including possible massless and partially
massless limits. The last section devoted to analogous construction
for the tensor $R_{\mu\nu}{}^{ab}$ having the symmetries of Riemann
tensor.

\section{Second rank tensor}

Let us start with the simple example of second rank tensor 
$h_\mu{}^a$. Strictly speaking it is not a mixed symmetry tensor, but 
this case turns out to be very instructive and interesting by itself. 
To have gauge invariant description of massive spin-2 particle one has 
to introduce two additional (Goldstone) fields --- vector $A_\mu$ and
scalar $\varphi$ ones. To be consistent in what follows we will use
first order formalism for all massless components which will serve
as building blocks for our massive particles. So we introduce three
pairs of fields: $(\omega_\mu{}^{ab}, h_\mu{}^a)$,
$(F^{ab}, A_\mu)$ and $(\pi^a, \varphi)$. As in our previous papers
on this subject \cite{Zin01,Zin02} our starting point will be just
the sum of the massless Lagrangians in flat Minkowski space:
\begin{eqnarray}
{\cal L}_0 &=& {\cal L}_0 (\omega_\mu{}^{ab}, h_\mu{}^a) +
{\cal L}_0 (F^{ab}, A_\mu) + {\cal L}_0 (\pi^a, \varphi) \\
{\cal L}_0 (\omega_\mu{}^{ab}, h_\mu{}^a) &=& \frac{1}{2}
\left\{ \phantom{|}^{\mu\nu}_{ab}
 \right\} \omega_\mu{}^{ac} \omega_\nu{}^{bc} - \frac{1}{2}
\left\{ \phantom{|}^{\mu\nu\alpha}_{abc} \right\}
\omega_\mu{}^{ab} \partial_\nu h_\alpha{}^c \nonumber \\
{\cal L}_0 (F^{ab}, A_\mu) &=& \frac{1}{4} F_{ab}{}^2 -
\frac{1}{2} \left\{ \phantom{|}^{\mu\nu}_{ab}  \right\}
F^{ab} \partial_\mu A_\nu \nonumber \\
{\cal L}_0 (\pi^a, \varphi) &=& - \frac{1}{2} \pi_a{}^2 +
\left\{ \phantom{|}^\mu_a \right\} \pi^a \partial_\mu \varphi
\nonumber
\end{eqnarray} 
Here
$$
\left\{ \phantom{|}^{\mu\nu}_{ab} \right\} = \delta_a{}^\mu
\delta_b{}^\nu - \delta_a{}^\nu \delta_b{}^\mu
$$
and so on. This Lagrangian is invariant under the following local
gauge transformations:
\begin{equation}
\delta_0 h_{\mu a} = \partial_\mu \xi_a + \eta_{\mu a} \qquad
\delta_0 \omega_\mu{}^{ab} = \partial_\mu \eta^{ab} \qquad
\delta_0 A_\mu = \partial_\mu \Lambda
\end{equation} 

Working with the first order formalism it is very convenient to 
use tetrad formulation of the underlying (Anti) de Sitter space.
We denote tetrad as $e_\mu{}^a$ (let us stress that it is not a
dynamical quantity here, just a background field) and Lorentz
covariant derivative as $D_\mu$. (Anti) de Sitter space is a 
constant curvature space with zero torsion, so we have:
\begin{equation}
D_{[\mu} e_{\nu]}{}^a = 0, \qquad [ D_\mu, D_\nu ] v^a =
R_{\mu\nu}{}^{ab} v_b = \kappa (e_\mu{}^a e_\nu{}^b -
e_\mu{}^b e_\nu{}^a) v_b
\end{equation} 
where $\kappa = - 2 \Lambda/(d-1)(d-2)$.

Now we replace all the derivatives in the Lagrangian and gauge
transformation laws by the covariant ones. Due to noncommutativity
of covariant derivatives the Lagrangian becomes not invariant
and we get:
\begin{equation}
\delta_0 {\cal L}_0 = \kappa (d-2) \omega^a \xi_a - \kappa
(d-2) h^{ab} \eta_{ab}
\end{equation} 

So we proceed by adding to the Lagrangian additional low derivatives
terms:
\begin{equation}
{\cal L}_1 = a_1 \left\{ \phantom{|}^{\mu\nu}_{ab} \right\}
\omega_\mu{}^{ab} A_\nu + a_2 \left\{ \phantom{|}^\mu_a \right\}
F^{ab} h_\mu{}^b + a_3 \left\{ \phantom{|}^\mu_a \right\}
\pi^a A_\mu
\end{equation} 
as well as corresponding corrections to the gauge transformation laws:
\begin{equation}
\delta_1 h_\mu{}^a = \alpha_1 e_\mu{}^a \Lambda \qquad
\delta_1 F^{ab} = \alpha_2 \eta^{ab} \qquad
\delta_1 A_\mu = \alpha_3 \xi_\mu \qquad
\delta_1 \varphi = \alpha_4 \Lambda
\end{equation} 
In this, the requirement that all variations in that order cancel
each other i.e. $\delta_0 {\cal L}_1 + \delta_1 {\cal L}_0 = 0$
gives us the following relations:
$$ 
\alpha_1 = \frac{2a_1}{d-2}, \quad \alpha_2 = - 2a_1, \quad
\alpha_3 = a_1, \quad \alpha_4 = - a_3, \quad a_2 = a_1
$$
At last, we introduce all possible mass-like terms into Lagrangian:
\begin{equation}
{\cal L}_2 = b_1 \left\{ \phantom{|}^{\mu\nu}_{ab} \right\}
h_\mu{}^a h_\nu{}^b + b_2 \left\{ \phantom{|}^\mu_a \right\}
h_\mu{}^a \varphi + b_3 \varphi^2
\end{equation} 
as well as necessary corrections to the gauge transformations:
\begin{equation}
\delta_2 \omega_\mu{}^{ab} = \beta_1 (e_\mu{}^a \xi^b - e_\mu{}^b 
\xi^a)\qquad \delta_2 \pi^a = \beta_2 \xi^a
\end{equation} 
Then, if one requires that all new variations
$\delta_0 {\cal L}_2 + \delta_1 {\cal L}_1 + \delta_2 {\cal L}_0$
cancel each other (taking into account the residue of
$\delta_0 {\cal L}_0$), then one could express all the parameters
in the Lagrangian and the gauge transformation laws in terms of
two parameters, say $a_1$ and $a_3$:
$$
\beta_1 = - \frac{2b_1}{d-2}, \quad 
\beta_2 = b_2 = a_1 a_3, \quad
b_1 = a_1{}^2 + \frac{\kappa(d-2)}{2}, \quad 
b_3 = \frac{d}{d-2} a_1{}^2
$$
In this, the whole Lagrangian will be invariant under all gauge
transformations provided:
\begin{equation}
4 (d-1) a_1{}^2 - (d-2) a_3{}^2 = - 2 \kappa (d-1)(d-2)
\end{equation} 

Now, having in our disposal complete description of general massive
particle in (A)dS space we can investigate all possible massless or
partially massless limits. Recall that parameter $\kappa$ is 
proportional to the cosmological constant $\Lambda$ so it could
be positive as well as negative. As a result one get two possible
special cases. In the de Sitter space $(\kappa < 0)$ one can set
$a_3 = 0$. In this, scalar component $(\pi^a, \varphi)$ completely
decouples from the system, while the rest fields with the Lagrangian:
\begin{equation}
{\cal L} = {\cal L}_0 (\omega_\mu{}^{ab}, h_\mu{}^a) +
{\cal L}_0 (F^{ab}, A_\mu) + 
m \left\{ \phantom{|}^{\mu\nu}_{ab} \right\} \omega_\mu{}^{ab} A_\nu +
m \left\{ \phantom{|}^\mu_a \right\} F^{ab} h_\mu{}^b
\end{equation} 
where $m^2 = - \kappa(d-2)/2$, which is invariant under the following
gauge transformations:
\begin{eqnarray}
\delta h_\mu{}^a &=& D_\mu \xi^a + e_{\mu b} \eta^{ba} +
\frac{2m}{d-2} e_\mu{}^a \Lambda \qquad
\delta \omega_\mu{}^{ab} = D_\mu \eta^{ab} \nonumber \\
\delta A_\mu &=& D_\mu \Lambda + m e_{\mu a} \xi^a 
\qquad \delta F^{ab} = - 2m \eta^{ab}
\end{eqnarray} 
describe the rather well known partially massless spin-2 particle.
Note, that in this particular limit there are no any explicit 
mass-like terms in the Lagrangian.

On the other side, in the Anti ds Sitter space one can set $a_1 = 0$.
In this, the whole system also decompose onto two subsystems. One of
them is just a usual massless spin-2 particle with the Lagrangian:
\begin{equation}
{\cal L} = {\cal L}_0 (\omega_\mu{}^{ab}, h_\mu{}^a) +
\frac{\kappa(d-2)}{2} \left\{ \phantom{|}^{\mu\nu}_{ab} \right\}
h_\mu{}^a h_\nu{}^b
\end{equation} 
and corresponding gauge transformations:
\begin{equation}
\delta h_\mu{}^a = D_\mu \xi^a + e_{\mu b} \eta^{ba} \qquad
\delta \omega_\mu{}^{ab} = D_\mu \eta^{ab} -
\kappa (e_\mu{}^a \xi^b - e_\mu{}^b \xi^a)
\end{equation} 

The other subsystem with the Lagrangian:
\begin{equation}
{\cal L} = {\cal L}_0 (F^{ab}, A_\mu) + {\cal L}_0 (\pi^a, \varphi)
 + m \left\{ \phantom{|}^\mu_a \right\} \pi^a A_\mu
\end{equation} 
where $ m^2 = 2 \kappa(d-1) $, which is invariant under
$$ 
\delta A_\mu = D_\mu \Lambda \qquad \delta \varphi = - m \Lambda
$$
describes massive vector particle (in a gauge invariant first order
formalism). Note that in such description there is no explicit
mass term for the vector field $A_\mu$, but if one solves (algebraic)
equation of motion for the field $\pi^a$ and puts the result back
into the Lagrangian one obtains appropriate term.

\section{$\Phi_{[\mu\nu],\alpha}$ tensor}

In this section we consider a truly mixed symmetry tensor
$\Phi_{[\mu\nu],\alpha}$. In \cite{Zin02} we have shown that gauge
invariant description of corresponding massive particle requires
introduction of three additional Goldstone fields: second rank
tensor $h_\mu{}^a$, two form $B_{[\mu\nu]}$ and vector $A_\mu$ ones.
To construct appropriate first order formalism we introduce four pairs
of fields: $(\Omega_\mu{}^{abc}, \Phi_{\mu\nu}{}^a)$,
$(\omega_\mu{}^{ab}, h_\mu{}^a)$, $(C^{abc}, B_{\mu\nu})$ and
$(F^{ab}, A_\mu)$. We start with the sum of massless first order
Lagrangians in flat Minkowski space:
\begin{eqnarray}
&& {\cal L}_0 = {\cal L}_0 (\Omega_\mu{}^{abc}, \Phi_{\mu\nu}{}^a) +
{\cal L}_0 (\omega_\mu{}^{ab}, h_\mu{}^a) +
{\cal L}_0 (C^{abc}, B_{\mu\nu}) + {\cal L}_0 (F^{ab}, A_\mu) \\
&& {\cal L}_0 (\Omega_\mu{}^{abc}, \Phi_{\mu\nu}{}^a) = - \frac{3}{4}
\left\{\phantom{|}^{\mu\nu}_{ab} \right\}
\Omega_\mu{}^{acd} \Omega_\nu{}^{bcd} + \frac{1}{4}
\left\{\phantom{|}^{\mu\nu\alpha\beta}_{abcd} \right\}
\Omega_\mu{}^{abc} \partial_\nu \Phi_{\alpha\beta}{}^d  \nonumber \\
&& {\cal L}_0 (C^{abc}, B_{\mu\nu}) = - \frac{1}{6} C_{abc}{}^2 +
\frac{1}{6} \left\{\phantom{|}^{\mu\nu\alpha}_{abc} \right\}
C^{abc} \partial_\mu B_{\nu\alpha} \nonumber
\end{eqnarray} 
where ${\cal L}_0 (\omega_\mu{}^{ab}, h_\mu{}^a)$ and
${\cal L}_0 (F^{ab}, A_\mu)$ are the same as in the previous section.
This Lagrangian is invariant under the following set of local gauge
transformations:
\begin{eqnarray}
\delta_0 \Phi_{\mu\nu}{}^a &=& \partial_\mu z_\nu{}^a - \partial_\nu
z_\mu{}^a + \eta_{\mu\nu}{}^a \qquad
\delta_0 \Omega_\mu{}^{abc} = \partial_\mu \eta^{abc} \qquad
\delta_0 B_{\mu\nu} = \partial_\mu \zeta_\nu - \partial_\nu \zeta_\mu
\nonumber \\
\delta_0 h_{\mu a} &=& \partial_\mu \xi_a + \eta_{\mu a} \qquad
\delta_0 \omega_\mu{}^{ab} = \partial_\mu \eta^{ab} \qquad
\delta_0 A_\mu = \partial_\mu \Lambda
\end{eqnarray} 

Now we replace all the derivatives in the Lagrangian and gauge
transformation laws by the Lorentz covariant ones (with the same
notations and conventions for the description of the (Anti) de Sitter
space as in the previous section). As usual, the Lagrangians
becomes to be non invariant under the gauge transformations and
we get:
\begin{equation}
\delta_0 {\cal L}_0 = 3 \kappa (d-3) [ \frac{1}{2} \Phi^{abc}
\eta_{abc} - \Omega^{ab} z_{ab} ] + \kappa (d-2) [ \omega^a \xi_a -
h^{ab} \eta_{ab} ]
\end{equation} 
Now we add to the Lagrangian all possible low derivatives terms
which could be written as forms with equal number of "world" and
"local" indices:
\begin{eqnarray}
{\cal L}_1 &=& a_1 \left\{\phantom{|}^{\mu\nu}_{ab} \right\}
\Omega_\mu{}^{abc} h_\nu{}^c + a_2 
\left\{\phantom{|}^{\mu\nu\alpha}_{abc} \right\}
\omega_\mu{}^{ab} \Phi_{\nu\alpha}{}^c + a_3
\left\{\phantom{|}^{\mu\nu\alpha}_{abc} \right\} 
\Omega_\mu{}^{abc} B_{\nu\alpha} + a_4
\left\{\phantom{|}^{\mu\nu}_{ab} \right\} C^{abc} \Phi_{\mu\nu}{}^c
 + \nonumber \\
 && + a_5 \left\{\phantom{|}^{\mu\nu}_{ab} \right\}
\omega_\mu{}^{ab} A_\nu + a_6 \left\{\phantom{|}^\mu_a \right\}
F^{ab} h_\mu{}^b + a_7 \left\{\phantom{|}^{\mu\nu}_{ab} \right\}
F^{ab} B_{\mu\nu} 
\end{eqnarray} 
as well as corresponding terms to the gauge transformation laws:
\begin{eqnarray}
\delta_1 \Phi_{\mu\nu} &=& \alpha_1 ( e_\mu{}^a \xi^b - e_\mu{}^b 
\xi^a)+ \alpha_2 (e_\mu{}^a \zeta^b - e_\mu{}^b \zeta^a) \qquad
\delta_1 \Omega_\mu{}^{abc} = \alpha_3 e_\mu{}^{[a} \eta^{bc]} 
\nonumber \\\delta_1 h_\mu{}^a &=& \alpha_4 z_\mu{}^a + \alpha_5 
e_\mu{}^a \Lambda\qquad \delta_1 \omega_\mu{}^{ab} = \alpha_6 e_{\mu 
c} \eta^{cab}  \\\delta_1 B_{\mu\nu} &=& \alpha_7 z_{\mu\nu} \qquad
\delta_1 C^{abc} = \alpha_8 \eta^{abc} \qquad
\delta_1 F^{ab} = \alpha_9 \eta^{ab} \qquad
\delta_1 A_\mu = \alpha_{10} \xi_\mu + \alpha_{11} \zeta_\mu \nonumber
\end{eqnarray} 
Note that in this case we can't add to the Lagrangian any explicit
mass-like terms, because the only possible term
$\left\{ \phantom{|}^{\mu\nu}_{ab} \right\} h_\mu{}^a h_\nu{}^b$
is forbidden by $\Lambda$-symmetry. Nevertheless it turned out
possible to achieve full gauge invariance of the Lagrangian by
appropriate choice of the parameters. Once again all the
parameters could be expressed in terms of two ones (we choose
$a_1$ and $a_3$)
$$ 
\alpha_1 = \frac{2a_1}{3(d-3)}, \quad \alpha_2 = \frac{4a_3}{d-3},
\quad \alpha_3 = 2\alpha_1, \quad \alpha_4 = 4a_1, \quad
\alpha_5 = \frac{2a_5}{d-2}, \quad \alpha_6 = - 2a_1
$$

$$ 
\alpha_7 = - 4 a_3, \quad \alpha_8 = 6 a_3, \quad \alpha_9 = - 2 a_5,
\quad \alpha_{10} = a_5, \quad \alpha_{11} = 4 a_7
$$

$$ 
a_2 = a_1, \quad a_4 = a_3, \quad a_6 = a_5, \quad
a_5 = 2 \sqrt{\frac{d-2}{d-3}} a_3, \quad 
a_7 = \sqrt{\frac{d-2}{d-3}} a_1
$$
In this, the following constraint must be satisfied:
\begin{equation}
a_1{}^2 - 3 a_3{}^2 = \frac{3}{8} \kappa (d-3)
\end{equation} 

In the gauge invariant description of massive particles we used to
work the massless limit is just the limit where all Goldstone
fields completely decouple from the main gauge field. For the case
at hand it means that one has to set $a_1 = 0$ and $a_3 = 0$
simultaneously. But the last relation clearly shows that it is
possible in the flat Minkowski space $\kappa = 0$ only. So there is
no truly massless limit for the field $\Phi_{\mu\nu}{}^a$ in
(A)dS space. Instead there exist two partially massless limits
depending on the sign of the cosmological term. In the Anti de Sitter
space one can set $a_3 = 0$. In this, the whole system decouples onto
two disconnected subsystems. One of them with the Lagrangian:
\begin{equation}
{\cal L} = {\cal L}_0 (\Omega_\mu{}^{abc}, \Phi_{\mu\nu}{}^a) +
{\cal L}_0 (\omega_\mu{}^{ab}, h_\mu{}^a) + 
m \left\{\phantom{|}^{\mu\nu}_{ab} \right\}
\Omega_\mu{}^{abc} h_\nu{}^c + m
\left\{\phantom{|}^{\mu\nu\alpha}_{abc} \right\}
\omega_\mu{}^{ab} \Phi_{\nu\alpha}{}^c
\end{equation} 
where $ m^2 = \frac{3}{8} \kappa (d-3) $ is invariant under the
following set of gauge transformations:
\begin{eqnarray}
\delta \Phi_{\mu\nu}{}^a &=& D_\mu z_\nu{}^a - D_\nu z_\mu{}^a + 
\eta_{\mu\nu}{}^a + \frac{2m}{3(d-3)} ( e_\mu{}^a \xi_\nu -
e_\nu{}^a \xi_\mu ) \nonumber \\
\delta \Omega_\mu{}^{abc} &=& 
D_\mu \eta^{abc} + \frac{4m}{3(d-3)} e_\mu{}^{[a} \eta^{bc]} \\
\delta h_\mu{}^a &=& D_\mu \xi^a + \eta_\mu{}^a + 4m z_\mu{}^a 
\qquad \delta \omega_\mu{}^{ab} = D_\mu \eta^{ab} - 2m \eta_\mu{}^{ab}
\nonumber
\end{eqnarray} 
and describes partially massless theory \cite{BMV00}. The rest fields
with the Lagrangian:
\begin{equation}
{\cal L} = {\cal L}_0 (C^{abc}, B_{\mu\nu}) + {\cal L}_0 (F^{ab}, 
A_\mu)+ \frac{M}{4} \left\{\phantom{|}^{\mu\nu}_{ab} \right\} F^{ab} 
B_{\mu\nu} \end{equation} 
where $ M^2 = 6 \kappa (d-2) $ and gauge transformations:
\begin{equation}
\delta B_{\mu\nu} = D_\mu \zeta_\nu - D_\nu \zeta_\mu \qquad
\delta A_\mu = D_\mu \Lambda + M \zeta_\mu
\end{equation} 
gives a gauge invariant description of massive antisymmetric tensor
in (A)dS space. Note again the absence of explicit mass term.

On the other hand, in the de Sitter space one can set $a_1 = 0$.
Then one also obtains two decoupled subsystems. This time  our
mixed symmetry tensor $\Phi_{\mu\nu}{}^a$ combines with the two
form $B_{\mu\nu}$ and gives us another example of partially
massless theory with the Lagrangian:
\begin{equation}
{\cal L} = {\cal L}_0 (\Omega_\mu{}^{abc}, \Phi_{\mu\nu}{}^a) +
{\cal L}_0 (C^{abc}, B_{\mu\nu}) + m
\left\{\phantom{|}^{\mu\nu\alpha}_{abc} \right\} 
\Omega_\mu{}^{abc} B_{\nu\alpha} + m
\left\{\phantom{|}^{\mu\nu}_{ab} \right\} C^{abc} \Phi_{\mu\nu}{}^c
\end{equation} 
where $ m^2 = - \kappa(d-3)/8 $ and the following set of gauge
transformations:
\begin{eqnarray}
\delta \Phi_{\mu\nu}{}^a &=& D_\mu z_\nu{}^a - D_\nu z_\mu{}^a +
\eta_{\mu\nu}{}^a + \frac{4m}{d-3} ( e_\mu{}^a \zeta_\nu -
e_\nu{}^a \zeta_\mu ) \qquad 
\delta \Omega_\mu{}^{abc} = D_\mu \eta^{abc} \nonumber \\
\delta B_{\mu\nu} &=& D_\mu \zeta_\nu - D_\nu \zeta_\mu - 4m 
z_{\mu\nu}\qquad \delta C^{abc} = 6m \eta^{abc}
\end{eqnarray} 

In this, the rest fields $(h_\mu{}^a, A_\mu)$ gives exactly the same
partially massless spin-2 theory as in the previous section.

\section{$R_{[\mu\nu],[\alpha\beta]}$ tensor}

The results of the previous section could be easily generalized
to the case of the mixed tensors with arbitrary number of "world"
indices and the only "local" one $\Phi_{[\mu_1 ... \mu_n]}{}^a$.
We will not proceed along this line here. Instead, in this section
we consider more interesting field $R_{[\mu\nu],[\alpha\beta]}$
having the symmetry of Riemann tensor. As we have shown in 
\cite{Zin02} for gauge invariant description of appropriate massive 
particle one needs two additional Goldstone fields: $\Phi_{\mu\nu}{}^a$ 
and $h_\mu{}^a$. So to construct first order form of such description
we introduce three pairs of fields: $(\Sigma_{\mu\nu}{}^{abc},
R_{\mu\nu}{}^{ab})$, $(\Omega_\mu{}^{abc}, \Phi_{\mu\nu}{}^a)$
and $(\omega_\mu{}^{ab}, h_\mu{}^a)$. The sum of flat space massless
Lagrangians:
\begin{eqnarray}
{\cal L}_0 &=& {\cal L}_0 (\Sigma_{\mu\nu}{}^{abc}, R_{\mu\nu}{}^{ab})
+ {\cal L}_0 (\Omega_\mu{}^{abc}, \Phi_{\mu\nu}{}^a) +
{\cal L}_0 (\omega_\mu{}^{ab}, h_\mu{}^a) \\
{\cal L}_0 (\Sigma_{\mu\nu}{}^{abc}, R_{\mu\nu}{}^{ab})
&=& - \frac{3}{8} \left\{
\phantom{|}^{\mu\nu\alpha\beta}_{abcd} \right\}
\Sigma_{\mu\nu}{}^{abe} \Sigma_{\alpha\beta}{}^{cde} + \frac{1}{4}
\left\{ \phantom{|}^{\mu\nu\alpha\beta\gamma}_{abcde} \right\}
\Sigma_{\mu\nu}{}^{abc} \partial_\alpha R_{\beta\gamma}{}^{de}
\end{eqnarray}
where ${\cal L}_0 (\Omega_\mu{}^{abc}, \Phi_{\mu\nu}{}^a)$ and
${\cal L}_0 (\omega_\mu{}^{ab}, h_\mu{}^a)$ are the same as before
is invariant under the following gauge transformations:
\begin{eqnarray}
\delta_0 R_{\mu\nu}{}^{ab} &=& \partial_\mu \chi_\nu{}^{ab} -
\partial_\nu \chi_\mu{}^{ab} + \psi_{\mu,\nu}{}^{ab} -
\psi_{\nu,\mu}{}^{ab} \qquad \delta_0 \Sigma_{\mu\nu}{}^{abc} =
\partial_\mu \psi_\nu{}^{abc} - \partial_\nu \psi_\mu{}^{abc}
\nonumber \\
\delta_0 \Phi_{\mu\nu}{}^a &=& \partial_\mu z_\nu{}^a - \partial_\nu
z_\mu{}^a + \eta_{\mu\nu}{}^a \qquad
\delta_0 \Omega_\mu{}^{abc} = \partial_\mu \eta^{abc} \qquad
\delta_0 B_{\mu\nu} = \partial_\mu \zeta_\nu - \partial_\nu \zeta_\mu
\nonumber \\
\delta_0 h_{\mu a} &=& \partial_\mu \xi_a + \eta_{\mu a} \qquad
\delta_0 \omega_\mu{}^{ab} = \partial_\mu \eta^{ab}
\end{eqnarray} 

We start with the replacement of all derivatives in the Lagrangian
and gauge transformation laws by the covariant ones. As a result
Lagrangian ${\cal L}_0$ is not invariant now, instead we have:
\begin{eqnarray}
\delta_0 {\cal L}_0 &=& - 6\kappa (d-4) [ 2 \Sigma^{a,bc} \chi_{b,ca}
- \Sigma^a \chi_a + R^{ab,cd} \psi_{d,abc} + 2 R^{ab} \psi_{ab} ] +
\nonumber \\
 && + 3 \kappa (d-3) [ \frac{1}{2} \Phi^{abc}
\eta_{abc} - \Omega^{ab} z_{ab} ] + \kappa (d-2) [ \omega^a \xi_a -
h^{ab} \eta_{ab} ]
\end{eqnarray} 
To proceed we add all possible low derivative terms to the Lagrangian
(by possible we mean those that could be written as forms with
equal number of "world" and "local" indices):
\begin{equation}
{\cal L}_1 = a_1 \left\{ \phantom{|}^{\mu\nu\alpha\beta}_{abcd}
\right\} \Sigma_{\mu\nu}{}^{abc} \Phi_{\alpha\beta}{}^d + a_2
\left\{ \phantom{|}^{\mu\nu\alpha}_{abc} \right\}
R_{\mu\nu}{}^{ad} \Omega_\alpha{}^{bcd} + a_3
\left\{ \phantom{|}^{\mu\nu}_{ab} \right\} \Omega_\mu{}^{abc} 
h_\nu{}^c+ a_4 \left\{ \phantom{|}^{\mu\nu\alpha}_{abc} \right\} 
\omega_\mu{}^{ab} \Phi_{\nu\alpha}{}^c 
\end{equation} 
as well as corresponding terms to the gauge transformations:
\begin{eqnarray}
\delta_1 R_{\mu\nu}{}^{ab} &=& \alpha_1 e_{[\mu}{}^{[a} 
z_{\nu]}{}^{b]}\qquad \delta_1 \Sigma_{\mu\nu}{}^{abc} = \alpha_2
e_{[\mu}{}^{[a} \eta_{\nu]}{}^{bc]} \nonumber \\
\delta_1 \Phi_{\mu\nu}{}^a &=& \alpha_3 (\chi_{\mu,\nu}{}^a -
\chi_{\nu,\mu}{}^a) + \alpha_4 e_{[\mu}{}^a \xi_{\nu]} \nonumber \\
\delta_1 \Omega_\mu{}^{abc} &=& \alpha_5 \psi_\mu{}^{abc} + \alpha_6
e_\mu{}^{[a} \eta^{bc]}  \\
\delta_1 h_\mu{}^a &=& \alpha_7 z_\mu{}^a \qquad
\delta_1 \omega_\mu{}^{ab} = \alpha_8 \eta_\mu{}^{ab} \nonumber
\end{eqnarray} 

At this order the gauge invariance $(\delta_0 {\cal L}_1 +
\delta_1 {\cal L}_0 = 0)$ gives a number of relations among the
parameters:
$$ 
\alpha_1 = - \frac{2a_1}{4}, \quad \alpha_2 = \frac{a_1}{4}, \quad
\alpha_3 = 4 a_1, \quad \alpha_5 = - 8 a_1
$$

$$ 
\alpha_6 = \frac{4a_3}{3(d-3)}, \quad \alpha_7 = 4 a_3, \quad
\alpha_8 = - 2 a_3, \quad a_2 = - 3 a_1, \quad a_4 = a_3
$$
but to achieve complete invariance we have to introduce a number of
mass-like terms in the Lagrangian (this time they do exist):
\begin{equation}
{\cal L}_2 = b_1 \left\{ \phantom{|}^{\mu\nu\alpha\beta}_{abcd} 
\right\}R_{\mu\nu}{}^{ab} R_{\alpha\beta}{}^{cd} + b_2
\left\{ \phantom{|}^{\mu\nu\alpha}_{abc} \right\} R_{\mu\nu}^{ab}
h_\alpha{}^c + b_3 \left\{ \phantom{|}^{\mu\nu}_{ab} \right\}
h_\mu{}^a h_\nu{}^b
\end{equation} 
as well as necessary corrections to the gauge transformations:
\begin{equation}
\delta_2 \Sigma_{\mu\nu}{}^{abc} = \beta_1 e_{[\mu}{}^{[a}
\chi_{\nu]}{}^{bc]} + \beta_2 e_{[\mu}{}^{[a} e_{\nu]}{}^b \xi^{c]}
\qquad \delta_2 \omega_\mu{}^{ab} = \beta_3 \chi_\mu{}^{ab} +
\beta_4 e_\mu{}^{[a} \xi^{b]}
\end{equation} 
In this, by adjusting the values of all parameters:
$$ 
\beta_1 = - \frac{8b_1}{3(d-4)}, \quad \beta_2 = 
\frac{2b_2}{3(d-3)(d-4)},\quad \beta_3 = 4 b_2, \quad \beta_4 = - 
\frac{2b_3}{d-2}$$

$$ 
b_1 = - 3 a_1{}^2 - \frac{3}{8} \kappa (d-4), \quad
b_2 = - 4 a_1 a_3, \quad b_3 = - \frac{4(d-2)}{3(d-3)} a_3{}^2 + 
\frac{d-2}{2} \kappa
$$
we obtain gauge invariant description of massive $R_{\mu\nu}{}^{ab}$
field in the (A)dS space, provided:
\begin{equation}
24 (d-3) a_1{}^2 - 8 (d-4) a_3{}^2 = - 3 \kappa (d-3)(d-4)
\end{equation} 

All these results look very similar to the ones obtained in the
previous section, but there is an essential difference which could
be traced to different number of fields and the presence of explicit
mass-like terms in the Lagrangian. This time we also have two
special limits depending on the sign of cosmological constant.
But now in the Anti de Sitter space by setting $a_1 = 0$ one
obtains truly massless theory for the tensor $R_{\mu\nu}{}^{ab}$
with rather simple Lagrangian:
\begin{equation}
{\cal L} = {\cal L}_0 (\Sigma_{\mu\nu}{}^{abc}, R_{\mu\nu}{}^{ab})
- \frac{3}{8} \kappa (d-4)
\left\{ \phantom{|}^{\mu\nu\alpha\beta}_{abcd} \right\}
R_{\mu\nu}{}^{ab} R_{\alpha\beta}{}^{cd}
\end{equation} 
which is invariant under the following gauge transformations:
\begin{equation}
\delta R_{\mu\nu}{}^{ab} = D_\mu \chi_\nu{}^{ab} -
D_\nu \chi_\mu{}^{ab} + \psi_{\mu,\nu}{}^{ab} -
\psi_{\nu,\mu}{}^{ab} \qquad
\delta \Sigma_{\mu\nu}{}^{abc} =
D_\mu \psi_\nu{}^{abc} - D_\nu \psi_\mu{}^{abc}
+ \kappa e_{[\mu}{}^{[a} \chi_{\nu]}{}^{bc]}
\end{equation} 
At the same time the rest fields completely decouples and describe
exactly the same partially massless theory as in the previous section.

On the other hand, in the de Sitter space one can set $a_3 = 0$.
This gives us one more example of partially massless theory
the Lagrangian being:
\begin{equation}
{\cal L} = {\cal L}_0 (\Sigma_{\mu\nu}{}^{abc}, R_{\mu\nu}{}^{ab})
+ {\cal L}_0 (\Omega_\mu{}^{abc}, \Phi_{\mu\nu}{}^a) +
m \left\{ \phantom{|}^{\mu\nu\alpha\beta}_{abcd}
\right\} \Sigma_{\mu\nu}{}^{abc} \Phi_{\alpha\beta}{}^d - 3m
\left\{ \phantom{|}^{\mu\nu\alpha}_{abc} \right\}
R_{\mu\nu}{}^{ad} \Omega_\alpha{}^{bcd}
\end{equation} 
with $ m^2 = \kappa (d-4)/8 $, which is invariant under:
\begin{eqnarray}
\delta R_{\mu\nu}{}^{ab} &=& D_\mu \chi_\nu{}^{ab} -
D_\nu \chi_\mu{}^{ab} + \psi_{\mu,\nu}{}^{ab} -
\psi_{\nu,\mu}{}^{ab} - \frac{2m}{d-4} e_{[\mu}{}^{[a} z_{\nu]}{}^{b]}
\nonumber \\
\delta \Sigma_{\mu\nu}{}^{abc} &=&
D_\mu \psi_\nu{}^{abc} - D_\nu \psi_\mu{}^{abc} +
\frac{m}{d-4} e_{[\mu}{}^{[a} \eta_{\nu]}{}^{bc]} \nonumber \\
\delta \Phi_{\mu\nu}{}^a &=& D_\mu z_\nu{}^a - D_\nu
z_\mu{}^a + \eta_{\mu\nu}{}^a + 4m (\chi_{\mu,\nu}{}^a -
\chi_{\nu,\mu}{}^a) \\
\delta \Omega_\mu{}^{abc} &=& D_\mu \eta^{abc} -
8m \psi_\mu{}^{abc} \nonumber
\end{eqnarray} 
One more time note that in the partially massless limit we obtain
relatively simple Lagrangian without any explicit mass-like terms.

\end{document}